\documentclass[aps,prl,twocolumn,superscriptaddress]{revtex4}
\usepackage{graphicx}
\usepackage{latexsym}
\usepackage{amssymb}
\usepackage{amsmath}
\usepackage{amsfonts}
\usepackage{bm}
\usepackage{multirow}
\usepackage{color}

\newcommand{\beq}{\begin{equation}}
\newcommand{\eeq}{\end{equation}}
\newcommand{\beqn}{\begin{eqnarray}}
\newcommand{\eeqn}{\end{eqnarray}}

\begin{document}

\title{Construction and Field Theory of Bosonic Symmetry Protected Topological states
\\ beyond Group Cohomology}

%\author{Cenke's group}

%\affiliation{Department of physics, University of California,
%Santa Barbara, CA 93106, USA}

\author{Zhen Bi}

\author{Cenke Xu}

\affiliation{Department of physics, University of California,
Santa Barbara, CA 93106, USA}

\begin{abstract}

We construct a series of bosonic symmetry protected topological
(BSPT) states beyond group cohomology classification using
``decorated defects" approach. This construction is based on
topological defects of ordinary Landau order parameters, decorated
with the bosonic short range entangled (BSRE) states in $(4k+3)d$
and $(4k+5)d$ space-time (with $k$ being nonnegative integers),
which do not need any symmetry. This approach not only gives these
BSPT states an intuitive physical picture, it also allows us to
derive the effective field theory for all these BSPT states beyond
group cohomology.

\end{abstract}


\maketitle

\section{1. Introduction}

Bosonic symmetry protected topological (BSPT) states are bosonic
analogues of fermionic quantum spin Hall insulator and topological
insulator, which have trivial bulk spectrum but nontrivial
boundary spectrum, as long as the system preserves certain
symmetry. There are roughly two types of BSPT states, their
mathematical difference is whether they can be classified and
described by group cohomology~\cite{wenspt,wenspt2} and
semiclassical nonlinear sigma model field theory~\cite{xuclass}.
For example, the well-known $E_8$ bosonic short range entangled
(BSRE) state~\footnote{In this paper we define short range
entangled state as systems with gapped and nondegenerate bulk
spectrum, namely it has no topological entanglement entropy. In
our definition, SRE states include SPT states as a
subset.}~\cite{e8,kitaev_talk2} in $2d$ space~\footnote{Throughout
the paper, the term ``space" always refers to the real physical
space. The phrase space-time will always be written explicitly.},
and its higher dimensional generalizations~\cite{xu6d} cannot be
classified by group cohomology.

Any nontrivial SPT state's boundary state cannot exist by itself,
as long as the system preserves the necessary symmetry. This means
that the boundary of a SPT state must be ``anomalous". The
relation between boundary anomaly and bulk SPT states has been
studied systematically in Ref.~\onlinecite{wenanomaly}. If a
nontrivial SRE state does not need any symmetry to protect its
boundary, then its boundary must have gravitational anomaly. The
$2d$ $p+ip$ topological superconductor, and the $2d$ $E_8$ state
both have chiral edge states, which lead to gravitational anomaly.
Analogues of $2d$ $E_8$ state can be found in all even spatial
dimensions. In every $(4k+2)d$ space (or equivalently $(4k+3)d$
space-time), there is a BSRE state with $\mathbb{Z}$
classification described by action~\cite{xu6d} \beqn
\mathcal{S}_{(4k+3)d} = \int \frac{i K^{IJ}}{4\pi} C^I \wedge d
C^J, \label{6d}\eeqn where $C^I$ is a $2k+1$ form antisymmetric
gauge field, and $K^{IJ}$ is the Cartan matrix of the $E_8$ group.
These states have bosonic $2k-$dimensional membrane excitations in
the bulk, and perturbative gravitational anomalies at the
boundary~\cite{xu6d,wittenanomaly}; In every $(4k+4)d$ space (or
equivalently $(4k+5)d$ space-time), there is a BSRE state with
$\mathbb{Z}_2$ classification described by action \beqn
\mathcal{S}_{(4k+5)d} = \int \frac{i K^{IJ}}{4\pi} B^I \wedge d
B^J, \label{5d}\eeqn where $B^I$ is a $2k+2$ form antisymmetric
gauge field, and $K^{IJ} = i\sigma^y$. This theory with $k=0$
($4d$ space) has been studied carefully in
Ref.~\onlinecite{mcgreevyswingle}, and it was demonstrated that
its $3d$ boundary is an ``all fermion" $3d$
QED~\cite{wangpottersenthil} which cannot be independently
realized in $3d$ space, and it has a global gravitational
anomaly~\cite{kongwen}.

As was pointed out by Ref.~\onlinecite{kapustin1,kapustin4}, the
state Eq.~\ref{5d} can also have a time-reversal symmetry. For
instance, this action is invariant under $Z_2^T: i \rightarrow -i,
(B^1, B^2) \rightarrow (B^2, B^1)$. But this state is also stable
if the time-reversal symmetry is broken. In this paper we will
only count this state as a BSRE state without any symmetry.

All these BSRE states in even spatial dimensions have their
descendant BSPT states in higher dimensions. All these descendant
BSPT states are also beyond the group cohomology classification.
Recently, a systematic mathematical formalism for BSRE and BSPT
states has been proposed in Ref.~\onlinecite{wenso}, which was
based on cohomology of $G \times SO(\infty)$, where $G$ is the
symmetry group, and $SO(\infty)$ is supposed to describe the
gravitational anomaly. The purpose of the current work is to give
a physical construction and field theory description of BSPT
states beyond the ordinary group cohomology classification. Our
results are summarized in Table~\ref{t1}.

\section{2. generalities}
%\subsection{A. Decorated Topological Defects}

We will view the BSRE states without any symmetry in even spatial
dimensions (Eq.~\ref{6d} and Eq.~\ref{5d}) as base states. Our
general strategy for constructing other beyond-Group-Cohomology
BSPT states, is to first break part or all of the symmetry by
condensing an ordinary Landau order parameter, then
proliferating/condensing the topological defects of the Landau
order parameter. The nontrivial BSPT state and the trivial state
are distinguished by the nature of the topological defects:
nontrivial BSPT states corresponds to the case where the defects
are decorated with the BSRE states in Eq.~\ref{6d} or
Eq.~\ref{5d}. The first example of such beyond-Group-Cohomology
BSPT state, which is protected by Time Reversal Symmetry
$\mathcal{T}$, was discovered in Ref.~\onlinecite{senthilashvin}.
This state can be constructed by proliferating
$\mathcal{T}$-breaking domain walls decorated with the $2d$ $E_8$
state. The topological term in the field theory Lagrangian density
that encodes the decoration reads:

\beqn
\mathcal{L}_{3+1d}^{Z_2^T}&=&\frac{i2\pi}{2\pi}ndn\wedge \frac{K^{IJ}_{E_8}}{8\pi^2}C^{I}\wedge dC^{J} \nonumber \\
&=&id\theta\wedge \frac{K^{IJ}_{E_8}}{8\pi^2}C^{I}\wedge dC^{J}\nonumber \\
&=&-i\theta \frac{K^{IJ}_{E_8}}{8\pi^2}dC^{I}\wedge dC^{J}
\label{3dT}
\eeqn
where the $O(2)$ vector $\vec{n}$ is
parametrized as $\vec{n}=(\cos\theta,\sin\theta)$. The
$\mathcal{T}$-symmetry transformation is \beqn
Z_2^T:&& (n_1,n_2)\rightarrow(n_1,-n_2),\nonumber \\
&& \theta\rightarrow -\theta \eeqn One can verify that the
Eq.~\ref{3dT} is time-reversal invariant. Also, if we keep
time-reversal invariance, then $\langle n_2 \rangle = 0$, namely
$\langle \theta \rangle = 0$ or $\pi$, which precisely corresponds
to the trivial and nontrivial BSPT state discussed in
Ref.~\onlinecite{senthilashvin}. Meanwhile, across a
$\mathcal{T}$-breaking domain wall, $\theta$ continuously changes
from $-\pi+0^+$ to $\pi+0^-$. After integrating over the normal
direction, the effective field theory left on the domain wall
precisely describes a $2d$ $E_8$ state.

The idea of ``decorated domain wall" construction of SPT states
was further explored in Ref.~\onlinecite{chenluashvin}. Domain
wall of $Z_2$ or time-reversal symmetry is the simplest kind of
topological defect. In our current work we will construct
beyond-group-cohomology BSPT states using more general topological
defects of other symmetry groups. Here we want to clarify that in
our current work the concept ``topological defect" refers to a
topologically stable configurations of Landau order parameter
$\vec{n}$ in $d-$dimensional space $\mathbb{R}^d$ with a
singularity $\mathcal{I}$, and the singularity can be viewed as
the boundary of $\mathbb{R}^d - \mathcal{I}$. The Landau order
parameter $\vec{n}$ has a soliton configuration on $\mathbb{R}^d -
\mathcal{I}$, which has no singularity any more. For example, in
$2d$ space a vortex core is a singularity at the origin $(0,0)$,
and it can be viewed as the boundary of $\mathbb{R}^2 - (0 ,0) $,
which is topologically equivalent to a ring $S^1$. A vortex
configuration corresponds to a $1d$ soliton on $S^1$, based on the
simple fact $\pi_1[S^1]= \mathbb{Z}$. In $3d$ space a hedgehog
monopole core is again a singularity at $(0,0,0)$, and a hedgehog
monopole corresponds to a soliton on space $\mathbb{R}^3 - (0 ,0,
0) $, based on the fact $\pi_2[S^2]= \mathbb{Z}$.

In general, the field theories we will discuss in this work is a
combination of the $\Theta$-term of $\vec{n}$ discussed in
Ref.~\onlinecite{xuclass} and Chern-Simons form of $C^I$ or $B^I$
in Eq.~\ref{6d},\ref{5d}. The explicitly form of the topological
term in $D-$dimensional space-time is: \beq
\mathcal{L}_{Dd,A}=\frac{i\Theta}{\Omega_{D-(4k+4)}}n\underbrace{dn\wedge
...\wedge dn}_{D-(4k+4)}\wedge\frac{K^{IJ}}{8\pi^2}dC^I\wedge
dC^J, \label{eA} \eeq \beq
\mathcal{L}_{Dd,B}=\frac{i\Theta}{\Omega_{D-(4k+6)}}n\underbrace{dn\wedge
...\wedge
dn}_{D-(4k+6)}\wedge\frac{(i\sigma^y)^{IJ}}{8\pi^2}dB^I\wedge
dB^J, \label{eB} \eeq where $\vec{n}$ is a Landau order parameter
with a unit length. $\Omega_D=V_D\times D!$, $V_D$ is the volume
of the unit $D$-dimensional sphere. Here we assume all components
of order parameter $\vec{n}$ transform nontrivially under the
symmetry group.

The equations above are also effectively equivalent to the two
equations in the follows: \beq \mathcal{L}_{Dd,A} =
\frac{i\Theta}{\Omega_{D-(4k+3)}} n\underbrace{dn \wedge ...\wedge
dn}_{D-(4k+3)}\wedge\frac{K^{IJ}}{8\pi^2} C^I\wedge dC^J.
\label{eA2} \eeq \beq \mathcal{L}_{Dd,B}=
\frac{i\Theta}{\Omega_{D-(4k+5)}} n\underbrace{dn\wedge ...\wedge
dn}_{D-(4k+5)}\wedge\frac{(i\sigma^y)^{IJ}}{8\pi^2}B^I\wedge dB^J
\label{eB2}, \eeq where the component $n_1 $ does not transform
under any symmetry group, but the rest of the components all
transform nontrivially. The equivalence between the two
descriptions above can be made explicit by parametrizing $\vec{n}$
as: $\vec{n} = \left( \cos \theta, \sin \theta n_2, \sin \theta
n_3, \cdots \right) $, then following the derivation in
Eq.~\ref{3dT}, because the desired BSPT state is fully symmetric,
$\langle \theta \rangle$ must be either $0$ or $\pi$, which
corresponds to the trivial state and nontrivial BSPT state
respectively. And with $\langle \theta \rangle = \pi$,
Eq.~\ref{eA2},\ref{eB2} return to Eq.~\ref{eA} and \ref{eB}.

All the terms above are ``topological" in the sense that they are
invariant under local coordinate transformation, because they do
not involve the metric. We only wrote down the most important
topological terms explicitly, but the readers should be reminded
that there are other terms that guarantee the system is in a fully
gapped and nondegenerate phase. For example, we need a term $1/g
(\partial_\mu \vec{n})^2$ in the field theory to control the
dynamics of $\vec{n}$, and we must keep $g$ large enough to
disorder $\vec{n}$; we also need a BF theory
term\cite{senthilashvin} $\sim (d \mathcal{B})^2 +
\frac{1}{2\pi}\mathcal{B} \wedge dC $ to gap out all the
excitations of the $C^I$ field.

Naively, we can also write down the following field theory, with
all $\vec{n}$ components transforming non-trivially under
symmetry: \beqn \mathcal{L} = \frac{i\Theta}{\Omega}ndn\wedge
...\wedge dn\wedge\frac{K^{IJ}}{8\pi^2}C^I\wedge dC^J. \nonumber
\eeqn For example, we can write down such field theory in $3+1d$
space-time, with $\vec{n}$ being an O(2) vector, and $C^I$ a one
form vector gauge field. Then the physical meaning of this field
theory is that, the vortex core of $\vec{n}$ will host the
boundary state of the $2d$ $E_8$ state, which must be gapless.
Then this means that we can never achieve a fully gapped
nondegenerate state by proliferating the vortex loops. Thus this
field theory will always be gapless, unless we explicitly break
the $U(1)$ symmetry of $\vec{n}$. Therefore this field theory
describes the boundary of a $4d$ space, rather than a $3d$ bulk
state.

%\subsection{C. Rules of Classification}

For field theories in Eq. \ref{eA} and \ref{eB}, in general we
consider fixed points $\Theta=2\pi p$ with $p\in \mathbb{Z}$.
However, this does not mean that we have a $\mathbb{Z}$ classified
state. If we can show that two field theories, $\Theta=0$ and
$\Theta=2\pi q$ with certain $q\in\mathbb{Z}$, can be smoothly
connected without closing the bulk gap, then they must be in the
same phase. In that case, the classification will be reduced to
$\mathbb{Z}_q$.

Because our field theory is constructed with order parameter
$\vec{n}$ and Chern-Simons form of $C^I$ or $B^I$, the
classification will depend on both sectors.

For pure $C\wedge dC$ theory, the classification is $\mathbb{Z}$,
because its boundary state has perturbative gravitational
anomaly~\cite{wittenanomaly,xu6d}, which has $\mathbb{Z}$
classification. Then the classification of the mixed field theory
of $\vec{n}$ and $C^I$ only depends on the $\vec{n}$ sector.

For instance, we can take Eq.~\ref{3dT} as an example. Take two
copies of the field theories and couple them to each other: \beqn
\mathcal{L}&=&\frac{i2\pi}{2\pi}n_{(1)}dn_{(1)}\wedge
\frac{K^{IJ}_{E_8}}{8\pi^2}C^{I}_{(1)}\wedge dC^{J}_{(1)}
+(1\rightarrow 2)\nonumber \\
&+& \beta n_{2,(1)}\cdot n_{2,(2)}+\lambda d C^{I}_{(1)} \wedge
\star d C^{I}_{(2)}, \eeqn where $\star$ is the Hodge star
operator. Now we fix $\lambda$ at a negative value, and tune
$\beta$ from negative to positive. With negative $\beta$,
effectively $\vec{n}_{(1)}$ and $\vec{n}_{(2)}$ will align with
each other, thus $n_{2,(1)} = n_{2,(2)}$, $C_{(1)} = C_{(2)}$,
then the two theories will ``constructively interfere" with each
other, and the final theory effectively has $\Theta = 4\pi$; with
positive $\beta$, effectively $n_{2,(1)} = - n_{2,(2)}$, $C_{(1)}
= C_{(2)}$, thus the two theories will ``destructively interfere"
with each other, and the final theory effectively has $\Theta =
0$. Because both theories are fully gapped and nondegenerate in
the bulk, tuning the coupling between them does not close the bulk
gap (as long as the coupling is not too strong to overcome the
bulk gap), thus the two effective coupled theories with $\Theta=0$
and $\Theta=4\pi$ are smoothly connected without going through a
bulk phase transition. Therefore the classification for the state
Eq.~\ref{3dT} is $\mathbb{Z}_2$.

By contrast, let us consider a $U(1)$ BSPT in $4d$ space with the
following field theory: \beq
\mathcal{L}^{U(1)}_{4+1d}=\frac{i2\pi}{2\pi}ndn\wedge
\frac{K^{IJ}_{E_8}}{8\pi^2}dC^{I}\wedge dC^{J}. \label{4du10} \eeq
The $U(1)$ symmetry acts as $U(1): (n_1+in_2)\rightarrow
e^{i\phi}(n_1+in_2)$. Imagine we have two copies of the theory,
the only $U(1)$ symmetry allowed coupling between these two
theories would be $\beta \vec{n}_{(1)} \cdot \vec{n}_{(2)}$. Then
for either sign of $\beta$, $i.e.$ for either $\vec{n}_{(1)} \sim
\vec{n}_{(2)}$ or $\vec{n}_{(1)} \sim - \vec{n}_{(2)}$, the final
effective theory always has $\Theta = 4\pi$ (simply because
$(-1)^2 = +1$). Thus there is no symmetry allowed coupling that
can continuously connect $\Theta = 4\pi$ to $\Theta = 0$.
Therefore the classification for this $U(1)$ BSPT state is
$\mathbb{Z}$.

For pure $B\wedge dB$ theory, the classification is
$\mathbb{Z}_2$~\cite{mcgreevyswingle}, therefore the
classification of the mixed state can only be $\mathbb{Z}_2$ or
trivial depending on the classification on the $\vec{n}$ sector.

%Generally, the classification is $\mathbb{Z}_{<n,2>}$ if we assume
%the classification of $\vec{n}$ is $\mathbb{Z}_n$.

\section{3. Examples of BSPT Beyond Group Cohomology}

In this section we study examples of beyond-group-cohomology BSPT
states with various symmetries up to $6+1$ space-time dimensions.
All these states are constructed with Landau order parameters and
the $2d$ $E_8$ state or the $4d$ BSRE state in Eq.~\ref{5d}. Our
results are summarized in TABLE \ref{t1}. Our results are mostly
consistent with results in Ref.~\onlinecite{wenso}, exceptions are
highlighted in red in the table.

\begin{center}
\begin{table}
    \begin{tabular}{|c|c|c|c|c|}
    \hline
    Symmetry & $3+1d$ & $4+1d$ & $5+1d$ & $6+1d$ \\ \hline
    $U(1)$&$0$&$\mathbb{Z}$&$0$&$\mathbb{Z}\times \mathbb{Z}_2$ \\ \hline
    $Z_2$&$0$&$\mathbb{Z}_2$&$\mathbb{Z}_2$&$\mathbb{Z}_2^2$ \\ \hline
    $Z_2^T$&$\mathbb{Z}_2$&$0$&$\mathbb{Z}_2^2$&$\mathbb{Z}_2$ \\ \hline
    $U(1)\rtimes Z_2^T$&$\mathbb{Z}_2$&$\mathbb{Z}$&$\mathbb{Z}_2^2$&$\mathbb{Z}_2^3$ {\color{red}$(\mathbb{Z}_2^4)$} \\ \hline
    $U(1)\times Z_2^T$&$\mathbb{Z}_2$&$0$&$\mathbb{Z}_2^3$&$\mathbb{Z}_2^2$ {\color{red}$(\mathbb{Z}_2^3)$} \\ \hline
    $U(1)\rtimes Z_2$&$0$&$\mathbb{Z}_2$&$\mathbb{Z}_2^2$&$\mathbb{Z}\times \mathbb{Z}_2^3$ \\ \hline
    $U(1)\times Z_2$&$0$&$\mathbb{Z}\times\mathbb{Z}_2$&$\mathbb{Z}_2$&$\mathbb{Z}\times\mathbb{Z}_2^4$ \\ \hline
    \end{tabular}
\caption{BSPT beyond Group Cohomology constructed from decorated
topological defects. Please note that the states within group
cohomology classification is not listed here. The case for
$U(1)\times Z_2$ symmetry was not discussed in
Ref.~\onlinecite{wenso}. Our results largely agree with
Ref.~\onlinecite{wenso}. The results in Ref.~\onlinecite{wenso}
that do not fully agree with ours are highlighted in red.}
    \label{t1}
\end{table}
\end{center}

\subsection{A. $U(1)$ Symmetry}

$\bullet$ In $4d$ space, there is a series of BSPT states with
$U(1)$ symmetry that is beyond the group cohomology, their field
theory is given by:
\beq
\mathcal{L}^{U(1)}_{4+1d}=\frac{i2\pi
k}{2\pi}ndn\wedge\frac{K^{IJ}_{E_8}}{8\pi^2}dC^{I}\wedge dC^{J},
\label{4du1}
\eeq
where $C^I$'s are rank-1 gauge field, and $k$
can take arbitrary integer value. Physically this state can be
viewed as decorating the $2\pi$ vortex of $U(1)$ order parameter
$\vec{n} = (n_1, n_2)$ (which is a $2d$ membrane in this
dimension) with the $E_8$ state, and then proliferating the
vortices. As we have shown in the previous section, this phase has
$\mathbb{Z}$ classification.

The $3+1d$ boundary of this state can be a superfluid phase with
spontaneously U(1) symmetry breaking, whose vortex line hosts the
edge states of the $2d$ $E_8$ state, $i.e.$ a chiral conformal
field theory with central charge $c = 8$. If we couple $\vec{n}$
to a $U(1)$ gauge field, then after we integrate out the gapped
matter field $\vec{n}$, the boundary of the system will have a
mixed $U(1)$-gravitational anomaly, namely the stress tensor of
the system is no longer conserved inside the $U(1)$ flux at the
boundary. A similar mixed gauge-gravity anomaly was also studied
in Ref.~\onlinecite{juvengu}.

$\bullet$ In $6+1d$ space-time, there are two root states for
$U(1)$ BSPT states beyond group cohomology, the first state is
described by the following field theory: \beq
\mathcal{L}_{6+1d,A}^{U(1)}=\frac{i2\pi k}{12\pi^2}ndn\wedge
dn\wedge dn\wedge \frac{K^{IJ}_{E_8}}{8\pi^2}dC^{I}\wedge dC^{J}
\eeq with \beqn
U(1): &&(n_1+in_2)\rightarrow e^{i\phi}(n_1+in_2),\nonumber \\
&&(n_3+in_4)\rightarrow e^{i\phi}(n_3+in_4). \eeqn This state has
$\mathbb{Z}$ classification. The state is constructed by
decorating the $E_8$ states on the intersection of two $U(1)$
vortices, and then proliferate the vortices (the two-vortex
intersection is now a $2d$ brane in $6d$ space).

The field theory of the second root phase is \beq
\mathcal{L}_{6+1d,B}^{U(1)} = \frac{i2\pi}{2\pi}ndn\wedge
\frac{(i\sigma^y)^{IJ}}{8\pi^2}dB^{I}\wedge dB^{J} \eeq where
\beq U(1): (n_1+in_2)\rightarrow e^{i\phi}(n_1+in_2) \eeq and
$B$'s are 2-form fields. The state has $\mathbb{Z}_2$
classification according to our rules. And physically this field
theory corresponds to decorating the $U(1)$ vortex with the $4d$
BSRE state in Eq.~\ref{5d}.

\subsection{B. $Z_2$ Symmetry}

$\bullet$ In $4+1d$ space-time, there is one nontrivial
beyond-cohomology BSPT state with $Z_2$ symmetry, and this state
is the descendant of the $U(1)$ beyond Group Cohomology state in
the same dimension in the sense that it can be obtained by
breaking the $U(1)$ symmetry to its subgroup $Z_2$ from
Eq.~\ref{4du1}: \beq
\mathcal{L}_{4+1d}^{Z_2}=\frac{i2\pi}{2\pi}ndn\wedge
\frac{K^{IJ}_{E_8}}{8\pi^2}dC^{I}\wedge dC^{J} \eeq with \beq Z_2:
(n_1,n_2)\rightarrow -(n_1,n_2) \eeq while the classification of
the state is now reduced to $\mathbb{Z}_2$ because the $n$-sector
is now $\mathbb{Z}_2$ classified.

$\bullet$ In $5+1d$ space-time, there is a $\mathbb{Z}_2$
classified new state which is not a descendant of any $U(1)$ state
discussed in the previous subsection. Physically this state is
constructed by decorating the $Z_2$ domain wall with $4d$ BSRE
state: \beqn \mathcal{L}_{5+1d}^{Z_2} &=&
\frac{i2\pi}{2\pi}ndn\wedge\frac{
(i\sigma^y)^{IJ}}{8\pi^2}B^{I}\wedge dB^{J}
\nonumber \\
&=& id\theta\wedge\frac{(i\sigma^y)^{IJ}}{8\pi^2}B^{I}\wedge dB^{J} \nonumber \\
&=& -i\theta\frac{(i\sigma^y)^{IJ}}{8\pi^2}dB^{I}\wedge dB^{J}
\eeqn Here we parametrize $\vec{n}$ as
$\vec{n}=(\cos\theta,\sin\theta)$. The symmetry transformation is:
\beqn
Z_2: &&(n_1,n_2)\rightarrow (n_1,-n_2)\nonumber \\
&& (B^1,B^2)\rightarrow(B^2,B^1) \nonumber \\
&& \theta\rightarrow-\theta. \eeqn Notice that $B^I$ must
transform nontrivially under $Z_2$ symmetry, in order to guarantee
that the field theory is $Z_2$ invariant. We can also choose a
different transformation for $B^I$: $Z_2: B \rightarrow \sigma^z
B$, but this transformation is equivalent to the previous after a
basis change. In a $Z_2$ invariant state, $\langle n_2 \rangle =
0$, $i.e.$ $\langle \theta \rangle = 0$ or $\pi$, which
corresponds to the trivial and BSPT state respectively.

$\bullet$ In $6+1d$ space-time, there are two root states, both of
which are descendants of $U(1)$ BSPT states, and both have
$\mathbb{Z}_2$ classification: \beq
\mathcal{L}_{6+1d,A}^{Z_2}=\frac{i2\pi }{12\pi^2}ndn\wedge
dn\wedge dn\wedge \frac{K^{IJ}_{E_8}}{8\pi^2}dC^{I}\wedge dC^{J}
\eeq with \beq Z_2: (n_1,n_2,n_3,n_4)\rightarrow
-(n_1,n_2,n_3,n_4). \eeq \beq \mathcal{L}_{6+1d,B}^{Z_2} =
\frac{i2\pi}{2\pi}ndn\wedge\frac{ (i\sigma^y)^{IJ}}{8\pi^2}
dB^{I}\wedge dB^{J} \eeq with \beq Z_2: (n_1,n_2)\rightarrow
-(n_1,n_2). \eeq

\subsection{C. $Z_2^T$ Symmetry}

$\bullet$ In $3+1d$ space-time, it is well-known that there is a
BSPT state beyond Group Cohomology~\cite{senthilashvin}. The state
can be understood by decorating $Z_2^T$ domain walls with the $2d$
$E_8$ state: \beqn
\mathcal{L}_{3+1d}^{Z_2^T}&=&\frac{i2\pi}{2\pi}ndn\wedge
\frac{K^{IJ}_{E_8}}{8\pi^2}C^{I}\wedge dC^{J} \nonumber \\
&=&-i\theta\frac{K^{IJ}_{E_8}}{8\pi^2}dC^{I}\wedge dC^{J} \eeqn
with \beqn Z_2^T:&& (n_1,n_2)\rightarrow(n_1,-n_2) \nonumber \\
&& \theta\rightarrow -\theta. \eeqn $\theta$ is defined as before,
$\langle \theta \rangle = 0$ and $\pi$ correspond to the trivial
and BSPT state respectively. This state has $\mathbb{Z}_2$
classification.

$\bullet$ In $5+1d$ space-time, there are two root states, both
have $\mathbb{Z}_2$ classification. The field theory for the first
state reads: \beq
\mathcal{L}_{5+1d,A}^{Z_2^T}=\frac{i2\pi}{8\pi}ndn\wedge
dn\wedge\frac{K^{IJ}_{E_8}}{8\pi^2}dC^{I}\wedge dC^{J}
\label{5dz2t} \eeq with \beq Z_2^T:
(n_1,n_2,n_3)\rightarrow-(n_1,n_2,n_3). \eeq The physical meaning
of this state is most transparent if we start with a system with
an enlarged SO(3)$\times Z_2^T$ symmetry, and $\vec{n}$ forms a
vector under the SO(3) symmetry. Then Eq.~\ref{5dz2t} can be
viewed as decoration of the hedgehog monopole of $\vec{n}$ with
the $2d $ $E_8$ state. Weakly breaking the SO(3) symmetry while
preserving the $Z_2^T$ symmetry does not change the nature of this
state. Alternatively, we can view the hedgehog monopole as the
intersection of three $Z_2^T$ domain walls.

The field theory for the second root state is \beqn
\mathcal{L}_{5+1d,B}^{Z_2^T} &=& \frac{i2\pi}{2\pi} ndn \wedge
\frac{ (i\sigma^y)^{IJ}}{8\pi^2}B^{I}\wedge dB^{J} \nonumber\\
&=& -i\theta\frac{(i\sigma^y)^{IJ}}{8\pi^2}dB^{I}\wedge dB^{J}
\eeqn with \beqn
Z_2^T:&& (n_1,n_2)\rightarrow(n_1,-n_2)\nonumber \\
&& \theta\rightarrow-\theta. \eeqn This state can be viewed as
decoration of $Z_2^T$ domain wall with the $4d$ BSRE state.

$\bullet$ In $6+1d$ space-time, there is one new state with
$\mathbb{Z}_2$ classification: \beq
\mathcal{L}_{6+1d}^{Z_2^T}=\frac{i2\pi}{2\pi}ndn\wedge\frac{
(i\sigma^y)^{IJ}}{8\pi^2} dB^{I}\wedge dB^{J} \eeq with \beqn
Z_2^T:&& (n_1,n_2)\rightarrow-(n_1,n_2)\nonumber\\
&& (B^1,B^2)\rightarrow(B^2,B^1). \eeqn The state is constructed
by decorating the vortex of $\vec{n}$ (or the intersection of two
$Z_2^T$ domain walls) with the $4d$ BSRE state.

\subsection{D. $U(1)\rtimes Z_2^T$ Symmetry}

$\bullet$ In $3+1d$ space-time, there is one nontrivial
beyond-cohomology BSPT state with $U(1)\rtimes Z_2^T$ symmetry,
but it is identical to the $Z_2^T$ state in the same dimension,
$U(1)$ symmetry simply acts trivially.

$\bullet$ In $4+1d$ space-time, there is one root state with
$\mathbb{Z}$ classification: \beq \mathcal{L}^{U(1)\rtimes
Z_2^T}_{4+1d}=\frac{i2\pi
k}{2\pi}ndn\wedge\frac{K^{IJ}_{E_8}}{8\pi^2}dC^{I}\wedge dC^{J}.
\eeq with \beqn
U(1):&& (n_1+in_2)\rightarrow e^{i\phi}(n_1+in_2)\nonumber \\
Z_2^T:&& (n_1,n_2)\rightarrow (n_1,-n_2).
\eeqn

$\bullet$ In $5+1d$ space-time, there are two root states, both
are identical to the $Z_2^T$ state in the same dimension with
trivial $U(1)$ symmetry transformation, and both are
$\mathbb{Z}_2$ classified.

$\bullet$ In $6+1d$ space-time, in Ref.~\onlinecite{wenso} there
are $four$ root states, all $\mathbb{Z}_2$ classified. However, we
can only find $three$ $\mathbb{Z}_2$ classified root states by our
construction.  The first one is identical to the $Z_2^T$ state in
$6+1d$. The other two root states are given by: \beq
\mathcal{L}_{6+1d,A}^{U(1)\rtimes Z_2^T}=\frac{i2\pi
}{12\pi^2}ndn\wedge dn\wedge dn\wedge
\frac{K^{IJ}_{E_8}}{8\pi^2}dC^{I}\wedge dC^{J} \label{6du1z2t}
\eeq with \beqn
U(1): &&(n_1+in_2)\rightarrow e^{i\phi}(n_1+in_2),\nonumber \\
Z_2^T:&& (n_1,n_2,n_3,n_4)\rightarrow(n_1,-n_2,-n_3,-n_4) \eeqn
and
\beq
\mathcal{L}_{6+1d,B}^{U(1)\rtimes
Z_2^T}=\frac{i2\pi}{2\pi}ndn\wedge\frac{(i\sigma^y)^{IJ}}{8\pi^2}dB^{I}\wedge
dB^{J} \eeq with \beqn
U(1):&& (n_1+in_2)\rightarrow e^{i\phi}(n_1+in_2),\nonumber \\
Z_2^T:&& (n_1,n_2)\rightarrow(n_1,-n_2)
\eeqn

We suspect the state we missed here is the mixed SPT state
described by $E^d(G)$ in Ref.~\onlinecite{wenso}.

\subsection{E. $U(1) \times Z_2^T$ Symmetry}

$\bullet$ In $3+1d$ space-time, there is a state identical to the
pure $Z_2^T$ state with trivial $U(1)$ symmetry transformation.

$\bullet$ In $5+1d$ space-time, we find three $\mathbb{Z}_2$
classified root states. Two of them are identical to the $Z_2^T$
states in $5+1d$ space-time, with trivial $U(1)$ symmetry
transformation. The third state is given by: \beq
\mathcal{L}_{5+1d}^{U(1)\times Z_2^T}=\frac{i2\pi}{8\pi}ndn\wedge
dn\wedge\frac{K^{IJ}_{E_8}}{8\pi^2}dC^{I}\wedge dC^{J} \eeq with
\beqn
U(1):&& (n_1+in_2)\rightarrow e^{i\phi}(n_1+in_2),\nonumber \\
Z_2^T:&& (n_1,n_2,n_3)\rightarrow(n_1,n_2,-n_3). \eeqn This state
can be viewed as decorating the $2d$ $E_8$ state on the
intersection of a $Z_2^T$ domain wall and a $U(1)$ vortex (it can
also be viewed as the hedgehog monopole of $\vec{n}$), then
proliferating both the domain walls and vortices.

$\bullet$ In $6+1d$ space-time, in Ref.~\onlinecite{wenso} there
are $three$ $\mathbb{Z}_2$ classified root states. However, using
our method we can only construct $two$ $\mathbb{Z}_2$ classified
root states. The first one is identical to the $Z_2^T$ state with
trivial $U(1)$ symmetry transformation. The other one is: \beq
\mathcal{L}_{6+1d}^{U(1)\times
Z_2^T}=\frac{i2\pi}{2\pi}ndn\wedge\frac{
(i\sigma^y)^{IJ}}{8\pi^2}dB^{I}\wedge dB^{J} \eeq with \beqn
U(1):&& (n_1+in_2)\rightarrow e^{i\phi}(n_1+in_2),\nonumber \\
Z_2^T:&& (n_1,n_2)\rightarrow - (n_1,n_2), \ \ B^{1 (2)}
\rightarrow B^{2 (1)}. \eeqn

One may ask whether field theory like Eq.~\ref{6du1z2t} could
correspond to a new root state. However, there is no consistent
way to assign the U(1)$\times Z_2^T$ symmetry transformations on
Eq.~\ref{6du1z2t}, namely Eq.~\ref{6du1z2t} cannot be invariant
under U(1)$\times Z_2^T$ symmetry, although it is invariant under
U(1)$\rtimes Z_2^T$ symmetry.

\subsection{F. $U(1)\rtimes Z_2=O_2$ Symmetry}

$\bullet$ In $4+1d$ space-time, there is one root state identical
to the BSPT state with $Z_2$ symmetry in the same dimension, the
$U(1)$ symmetry simply acts trivially.

$\bullet$ In $5+1d$ space-time, there are two root states, both
$\mathbb{Z}_2$ classified. One is the same $Z_2$ state with
trivial $U(1)$ action. The other one is given by: \beq
\mathcal{L}_{5+1d}^{U(1)\rtimes Z_2}=\frac{i2\pi}{8\pi}ndn\wedge
dn\wedge\frac{K^{IJ}_{E_8}}{8\pi^2}dC^{I}\wedge dC^{J} \eeq with
\beqn
U(1):&& (n_1+in_2)\rightarrow e^{i\phi}(n_1+in_2),\nonumber \\
Z_2:&& (n_1,n_2,n_3)\rightarrow(n_1,-n_2,-n_3). \eeqn This state
can be viewed as decorating the $2d$ $E_8$ state on the
intersection of $U(1)$ vortex and $Z_2$ domain wall. Also, the $O_2$
symmetry is a subgroup of $SO(3)$ symmetry, thus the vortex-domain
wall intersection is simply the hedgehog monopole of the $SO(3)$
vector $\vec{n}$.

$\bullet$ In $6+1d$ space-time, we find $four$ root states, which
is $more$ than the results in Ref.~\onlinecite{wenso}. Two of them
are the same as the BSPT states with $Z_2$ symmetry, both of which
are $\mathbb{Z}_2$ classified. The third root state is described
by \beq \mathcal{L}_{6+1d,A}^{U(1)\rtimes Z_2}=\frac{i2\pi k
}{12\pi^2}ndn\wedge dn\wedge dn\wedge
\frac{K^{IJ}_{E_8}}{8\pi^2}dC^{I}\wedge dC^{J} \eeq with \beqn
U(1): &&(n_1+in_2)\rightarrow e^{i\phi}(n_1+in_2),\nonumber \\
&& (n_3+in_4)\rightarrow e^{i\phi}(n_3+in_4),\nonumber \\
Z_2:&& (n_1,n_2,n_3,n_4)\rightarrow(n_1,-n_2,n_3,-n_4) \eeqn This
state is $\mathbb{Z}$ classified. This state can be viewed as
decorating the $2d$ $E_8$ state on the intersection of two
vortices, then proliferate the vortices afterwards.

The last root state in $ 6+1d$ space-time is described by \beq
\mathcal{L}_{6+1d,B}^{U(1)\rtimes
Z_2}=\frac{i2\pi}{2\pi}ndn\wedge\frac{
(i\sigma^y)^{IJ}}{8\pi^2}dB^{I}\wedge dB^{J} \eeq with \beqn
U(1):&& (n_1+in_2)\rightarrow e^{i\phi}(n_1+in_2),\nonumber \\
Z_2:&& (n_1,n_2)\rightarrow(n_1,-n_2),\nonumber\\
&& (B^1,B^2)\rightarrow (B^2,B^1).
\eeqn
This state has $\mathbb{Z}_2$ classification.

\subsection{G. $U(1)\times Z_2$ Symmetry}

$\bullet$ In $4+1d$ space-time, we have two root states, both of
which are descendants from pure $U(1)$ state and pure $Z_2$ state
respectively.

$\bullet$ In $5+1d$ space-time, there is only one root state,
which is the same as the state with $Z_2$ symmetry only.

$\bullet$ In $6+1d$ space-time, there are five root states. The
first three states can all be described by the same field theory:
\beq \mathcal{L}_{6+1d,A}^{U(1)\times Z_2}=\frac{i2\pi k
}{12\pi^2}ndn\wedge dn\wedge dn\wedge
\frac{K^{IJ}_{E_8}}{8\pi^2}dC^{I}\wedge dC^{J} \eeq These three
different states have the same form of Lagrangian, but they are
distinguished from each other by their symmetry transformations:
\beqn
(1)\ \  U(1):&& trivial, \nonumber \\
Z_2:&& (n_1,n_2,n_3,n_4)\rightarrow -(n_1,n_2,n_3,n_4).\\ \nonumber\\
(2)\ \  U(1):&& (n_1+in_2)\rightarrow e^{i\phi}(n_1+in_2), \nonumber \\
Z_2:&& (n_1,n_2,n_3,n_4)\rightarrow -(n_1,n_2,n_3,n_4).\\ \nonumber\\
(3)\ \  U(1):&& (n_1+in_2)\rightarrow e^{i\phi}(n_1+in_2), \nonumber \\
&& (n_3+in_4)\rightarrow e^{i\phi}(n_3+in_4), \nonumber \\
Z_2:&& (n_1,n_2,n_3,n_4)\rightarrow -(n_1,n_2,n_3,n_4). \eeqn The
classification of the three states are $\mathbb{Z}_2$,
$\mathbb{Z}_2$ and $\mathbb{Z}$ respectively.

The other two states are described by the following field theory:
\beq \mathcal{L}_{6+1d,B}^{U(1)\times
Z_2}=\frac{i2\pi}{2\pi}ndn\wedge\frac{(i\sigma^y)^{IJ}}{8\pi^2}dB^{I}\wedge
dB^{J} \eeq again, these two states have different transformations
under symmetry groups: \beqn
(4)\ \  U(1):&& trivial, \nonumber \\
Z_2:&& (n_1,n_2)\rightarrow -(n_1,n_2). \\
\nonumber \\
(5)\ \  U(1):&& (n_1+in_2)\rightarrow e^{i\phi}(n_1+in_2), \nonumber \\
Z_2:&& trivial.
\eeqn
The classification of the two states are both $\mathbb{Z}_2$.

\section{4. summary}

In this work, we construct field theories of
beyond-Group-Cohomology BSPT states based on decorated topological
defect picture. Our results are largely consistent with
Ref.~\onlinecite{wenso}, with a few exceptions. We listed examples
of BSPT states below six dimensional space, but our construction
can be straightforwardly generalized to all higher dimensions, as
long as we use the generalized base states in
Eq.~\ref{6d},\ref{5d} for $k \geq 1$. We also note that
Ref.~\onlinecite{juvengu} proposed the SPT states beyond group
cohomology should have mixed gauge-gravity responses, which is
also consistent with the formalism used in our current work.

The authors are supported by the the David and Lucile Packard
Foundation and NSF Grant No. DMR-1151208. The authors thank
Xiao-Gang Wen for many inspiring communications.

%\section{acknowledgement}

\bibliography{invertible}

\end{document}